\documentclass[OSA,JOSAB,twocolumn]{revtex4}

\usepackage{float}
\usepackage{amsmath}
\usepackage{amssymb}
\usepackage{graphicx}
\usepackage{changepage}
\usepackage{verbatim}

\begin{document}

\title{Testing the diffusion hypothesis as a mechanism of self-healing in Disperse orange 11 doped in PMMA}
\author{Shiva K. Ramini}
\affiliation{Department of Physics and Astronomy,Washington State University, Pullman, WA, 99163, USA}
\email{rshiva@wsu.edu}
\author{Nathan Dawson}
\affiliation{Department of Physics and Astronomy,Washington State University, Pullman, WA, 99163, USA}
\author{Mark G. Kuzyk}
\affiliation{Department of Physics and Astronomy,Washington State University, Pullman, WA, 99163, USA}
\date{\today}
\begin{abstract}
In this work, we show that reversible photodegradation of Disperse Orange 11 doped in PMMA is not due to dye diffusion -- a common phenomenon observed in many dye-doped polymers. The change in linear absorbance due to photodegradation of the material shows an isobestic point, which is consistent with the formation of a quasi-stable damaged species.  Spatially-resolved amplified spontaneous emission and fluorescence, both related to the population density, are measured by scanning the pump beam over a burn mark.  A numerical model of the time evolution of the population density due to diffusion is inconsistent with the experimental data suggesting that diffusion is not responsible.
\end{abstract}
\maketitle
\setcounter{tocdepth}{3}

\section{Introduction}
Photodegradation in dye doped polymers was extensively studied and modeled in several systems\cite{Sekkat1992,Galvan-Gonzalez2000a}.
Recently, reversible photodegradation was observed for dye molecules when doped in polymers\cite{howel02.01} but not in liquids\cite{howel04.01}. A logical hypothesis is that dye diffusion is responsible for this effect.

Dye diffusion is a common phenomenon in azo dye-doped polymer systems\cite{Galvan-Gonzalez2000a}. There are many applications that are based on dye diffusion such as thermal transfer printing\cite{Dubois1996} or holographic surface relief gratings\cite{kim1995laser}. Photoinduced dye diffusion is observed in various dye-polymer systems and has been thoroughly studied over the past decade\cite{Lefin1998,yager2001all}.

We previously reported on candidate mechanisms that could be responsible for reversible photodegradation\cite{embaye2008} in 1-amino-2-methyl anthraquinone (commonly known as Disperse orange 11) when doped in Poly(methyl methacrylate)(PMMA). This paper focuses specifically on eliminating the likelihood of dye diffusion alone as a cause of self-healing effects. Anderson \textit{et al}'s imaging studies\cite{Anderson2011} of photodamage show the visual recovery of a burn track, yet it alone does not fully rule out the possibility of diffusion.

In this paper, we present experimental observations of changes in absorption spectrum of a damaged area, which support the hypothesis of the formation of a damaged species and subsequent self healing back to the original molecule.  We also report on measurements of the time evolution of the population density profile and show that they are inconsistent with theoretically-simulated profiles due to diffusion.

\section{Experiments}
\begin{figure}
  \includegraphics{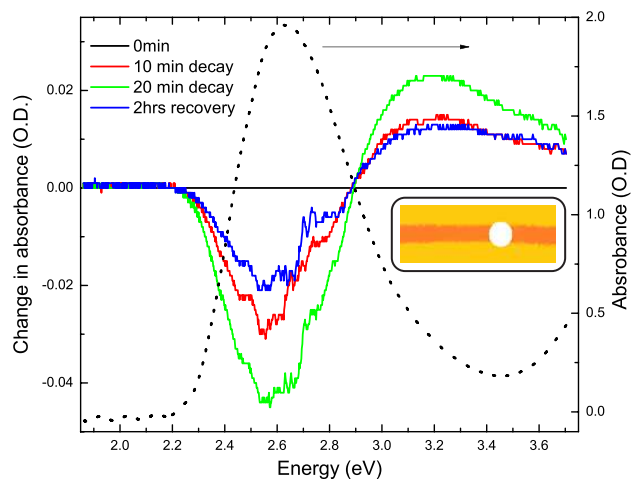}\\
  \caption{Isobestic behavior in linear absorbance suggests molecular structure changes. Sample used: 6g/l of disperse oragnge 11 doped in PMMA made into a thin film of thickness of the order of 100$\mu$m. Inset shows a cartoon of pump line (dark orange) and probe (white circle) overlapping in the sample.}
  \label{T090318Q-02}
\end{figure}

A frequency doubled Nd:YAG nanosecond pulsed laser at a wavelength of 532 nm is brought to a line focus of 100$\mu$m thick on a sample to induce optical damage.  This pump beam also produces amplified spontaneous emission (ASE) light that propagates along the excitation line.  The absorption spectrum is measured as a function of time using a white light probe that spatially overlaps the pump beam at the sample. The inset in Figure \ref{T090318Q-02} shows a cartoon of the overlap between the probe beam and pump beam line.

The sample is pumped for 20 minutes, during which time the absorption spectrum is measured at $t = 0$, $10$ and $20$ to probe photodegradation.  Subsequently, the laser is blocked to allow the sample to self heal. The absorption spectrum is measured again after 2 hours of healing.  The experiment is described in more detail in the literature\cite{embaye2008}.

Diffusion is characterized by a decrease of population in the bright regions, leading to an absorption spectrum of the same shape but decreased area.  Figure 1 shows the absorption spectrum plotted as a function of energy at various times with the initial spectrum subtracted. The dotted curve shows the initial absorption spectrum of the sample.

An isobestic point for a series of linear absorption spectra is an indication of the conversion of one molecular structure into another one, or changes in the molecular structure; but, is not expected for diffusion. The depletion of undamaged molecules is clear from the dip at 2.57eV that characterizes the DO11 molecule, while the growing peak at 3.2eV is an indication of the formation of a damaged population.  There is a clear isobestic point between these two regions.  After two hours, the damaged population characterized by the peak at 3.2eV decreases and the dip at 2.57eV decreases showing recovery of the sample. A two peak fit to the curves (not shown) approximately yields an equal area under the dip and peak showing the conservation of population of molecules in the region probed with the white light source.  While this data clearly shows the conversion between two molecular structures, it does not rule out contributions from diffusion.

The amplified spontaneous emission (ASE) intensity and the fluorescence intensity are related to the population density in the pump region.  As such, a mapping of the collected fluorescence or ASE light as a function of pump position can be used to determine the population profile as a function time.

\begin{figure}
  \includegraphics{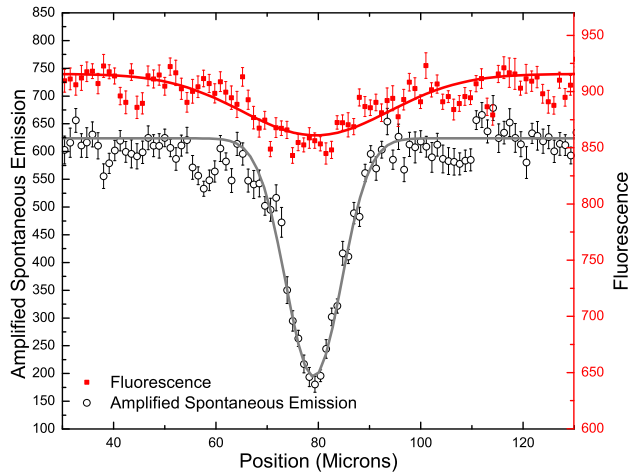}\\
  \caption{Fluorescence and ASE intensity as a function of excitation line position.}
  \label{110427-01}
\end{figure}
We use the line focus of the pump laser both to burn a damage line into the sample and to measure the population profile by monitoring fluorescence/ASE while translating the pump line through the burn line. Since the line focus is of transverse gaussian intensity profile, we expect the burn mark to also have a gaussian profile. The fluorescence/ASE profile will thus be a convolution of the undamaged population density profile and the pump intensity profile. If diffusion is a dominant mechanism, we would expect a buildup of population just outside the burn line due to photothermally-induced transport of molecules away from the hot area.  Figure \ref{110427-01} shows that there is no significant increase of fluorescence/ASE emission when the sample is pumped near the edges of the burn line.

\begin{figure}
  \includegraphics{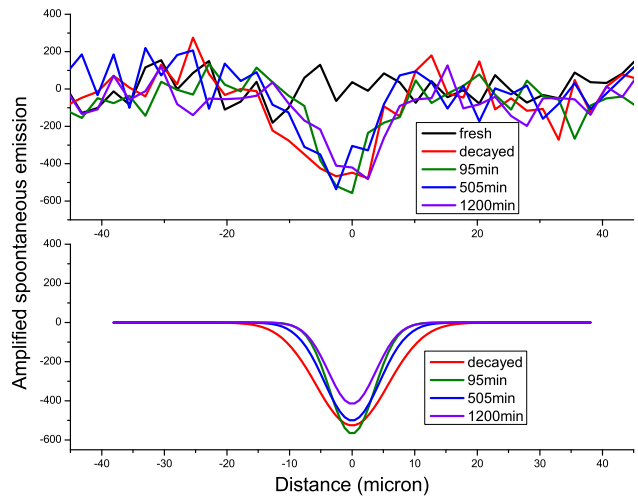}\\
  \caption{ASE scan over a burn mark as the sample recovers. Top: the experimental data at several times, and bottom: gaussian fits to the experimental data.}
  \label{110407-01}
\end{figure}

Since ASE is a nonlinear process, the width of the cross correlation function determined from a scan is narrower than it is for linear fluorescence -- as observed in Figure \ref{110427-01}. Figure \ref{110407-01} shows the time evolution of the ASE profile as a function of time during recovery. The decrease in the width of these curves is consistent with the fact that ASE is a nonlinear process.

\section{Theoretical modeling}

In this section, we present a model of dye diffusion in a polymer due to a photo-induced temperature gradient. We consider the contribution of diffusion to the mechanisms of photo-degradation and self-healing by simulating both the population loss in the illuminated region and the return of population when the sample recovers in the dark.

\subsection{Diffusion from a temperature gradient}

It is well known that the equilibrium population density of a molecular species is altered by a temperature gradient due to molecular diffusion in the host material\cite{Enge2004,lee1993effect}.  Fick's first law for particles in a material with a temperature gradient is
\begin{equation}
J = C M F\left(r,C\right) - D\left(r,C\right) \nabla C,
\label{eq:fickslaw}
\end{equation}
where $J$ is the diffusion flux, $C$ is the concentration, $M$ is the mobility, $F$ is the force from the temperature gradient, $D$ is the diffusivity, $t$ is the time, and $r$ is the position.

The force on dye molecules from the temperature gradient, $F$, can be expressed in terms of the temperature, $T$, and the transport heat, $Q$. This gives
\begin{equation}
F = -\frac{Q}{T}\nabla T .
\label{eq:transportforce}
\end{equation}
The mobility of the dye molecules in the polymer is given by
\begin{equation}
M = \frac{1}{C} D\left(r,C\right)\left(\frac{\partial \mu}{\partial C}\right)^{-1} ,
\label{eq:mobility}
\end{equation}
where $\mu$ is the chemical potential. Note that this assumes a constant volume. We now can rewrite Fick's first law for constant volume with concentration and temperature gradient as
\begin{equation}
J = - \left( C M \frac{Q}{T} \nabla T + D\left(r,C\right) \nabla C \right) .
\label{eq:dflux}
\end{equation}

The parameters used in each simulation are chosen according to the experimental conditions. The samples are approximately $100\,\mu$m thick and $1\,$cm square. The pump beam, of gaussian profile, is focused by a cylindrical lens onto the flat surface of the sample. The line focus spans the entire length of the sample in one direction. The transverse beam intensity profile is much smaller than the width of the sample.  Since measurements are taken in the center of the sample, we approximate the beam as an infinitely long heat source.

Next, we evaluate Fick's second law for a constant temperature gradient. The constant temperature gradient approximation due to a photo-induced heat source is valid for any system that exhibits a thermal diffusivity that is much greater than the diffusivity of the molecules through the medium, which is valid for a dye-doped polymer \cite{Li1997,Agari1997}. The diffusion equation for a material with a known temperature profile, $T(x,y)$, is
\begin{equation}
\frac{\partial C}{\partial t} = \nabla \left( C M \frac{Q}{T} \nabla T + D\left(r,C\right) \nabla C \right) .
\label{eq:diffT}
\end{equation}

\begin{figure}
\includegraphics[scale=1]{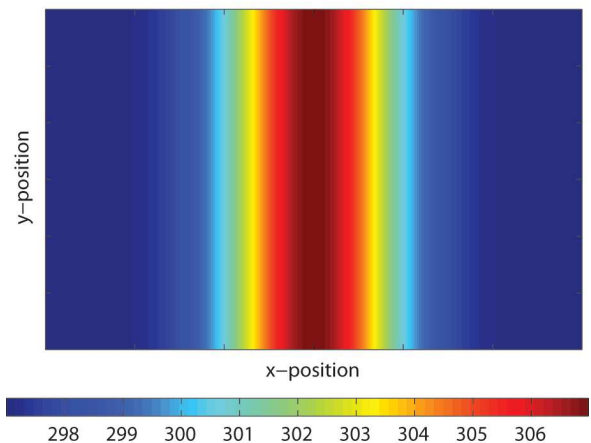}
\caption{The temperature profile from a pump beam that is infinitely long in the $y$-direction and has a gaussian intensity profile in the $x$-direction.}
\label{fig:templine}
\end{figure}
A uniform temperature distribution in the $y$-direction from an infinitely long heat source gives
\begin{equation}
\frac{d T}{d y} = 0 .
\label{eq:ytemp0}
\end{equation}
Figure \ref{fig:templine} shows the temperature profile throughout the material for an ambient temperature of $297\,$K and a peak temperature of $307\,$K, values that are typical for the pump powers used. Thus, for an initial uniform concentration
\begin{equation}
\frac{d C}{d y} = 0 .
\label{eq:yconcentration0}
\end{equation}
Therefore, the concentration only varies in the $x$-direction for a line focus heat source along $y$.

The diffusion equation in one dimension is
\begin{equation}
\frac{\partial C}{\partial t} = \frac{\partial}{\partial x} \left( C M \frac{Q}{T} \frac{\partial T}{\partial x} + D\left(x,C\right) \frac{\partial C}{\partial x} \right) .
\label{eq:diffTx}
\end{equation}
Assuming that the diffusivity is constant with respect to the small changes in concentration, Equation \ref{eq:diffTx} can be rewritten using the quotient rule in terms of separate derivatives of the concentration and temperature
\begin{eqnarray}
\frac{\partial C}{\partial t} &=& M Q \left[C\left(\frac{\partial}{\partial x}\frac{1}{T}\right) \left(\frac{\partial T}{\partial x}\right) + \frac{C}{T}\frac{\partial^2 T}{\partial x^2} \right. \nonumber \\
 &+& \left. \frac{1}{T}\left(\frac{\partial T}{\partial x}\right) \left(\frac{\partial C}{\partial x}\right)\right] + D \frac{\partial^2 C}{\partial x^2} ,
\label{eq:diffTxexplicit}
\end{eqnarray}
where $D$, $M$, and $Q$ are approximated as constants with respect to the concentration and the spatial coordinate. The temperature profile is
\begin{equation}
T = T_0+T_{\mathrm{max}}e^{-x^2/2\sigma^2} ,
\label{eq:tempprofilex}
\end{equation}
where Figure \ref{fig:templine} shows the temperature distribution in the plane of the thin-film sample.

Substituting Equation \ref{eq:tempprofilex} into Equation \ref{eq:diffTxexplicit}, and setting $T^\prime = T_{\mathrm{max}}/T_0$, gives
\begin{eqnarray}
\frac{\partial C}{\partial t} &=& D \frac{\partial^2 C}{\partial x^2} + M Q \displaystyle\frac{T^\prime e^{-x^2/2\sigma^2}}{1+T^\prime e^{-x^2/2\sigma^2}} \label{eq:diffTempxexp}\\
&\times& \left[ C \left( \frac{x^2}{\sigma^4} - \frac{x^2}{\sigma^4} \frac{T^\prime e^{-x^2/2\sigma^2}}{1+T^\prime e^{-x^2/2\sigma^2}} - \frac{1}{\sigma^2} \right) - \frac{x}{\sigma^2} \frac{\partial C}{\partial x}\right] \nonumber .
\end{eqnarray}
To avoid small errors at the boundaries due to numerical approximations, we impose Dirichlet (rather than Neumann) boundary conditions at $x=\pm L/2$, where $L$ is the length of each edge of the square sample and the boundaries are far from the heat source. This is especially important for computing small changes over long periods of time. The Dirichlet boundary conditions are
\begin{equation}
C\left(x=\pm \frac L 2,t\right) = C_0.
\end{equation}
We also assume that the concentration is uniform before the laser is turned on, or,
\begin{equation}
C\left(x,t=0\right) = C_0 .
\label{eq:initialconc}
\end{equation}

We begin by first solving the diffusion equation while the pump laser is on to determine the initial concentration profile when the laser is turned off and the sample begins to heal.  At this time, Equation \ref{eq:diffTempxexp} reduces to the simplest form of the diffusion equation,
\begin{equation}
\frac{\partial C}{\partial t} = D \frac{\partial^2 C}{\partial x^2} .
\label{eq:simplediffequation}
\end{equation}

Equation \ref{eq:diffTempxexp} has no transient solutions in an analytical form of common functions. Therefore, we must find a numerical approximation to determine the concentration as a function of time and position due to the temperature gradient induced by the laser. The Crank-Nicolson method is used to solve all diffusion equations.

We investigate two cases. The first involves pure diffusion, which assumes that there are no chemical reactions. The second assumes that there are two process; namely, a partial burn due to photochemistry and dye diffusion. For both cases, we use a nonlinear fitting scheme to approximately determine the parameters $M Q$, $D$, and $\sigma$ in the diffusion equation. Note that we assume $T^\prime = 10/297\,$K - values that are typical in our experiments. This is a good approximation because
\begin{equation}
\displaystyle\frac{T^\prime}{1+T^\prime e^{-x^2/2\sigma^2}} \approx T^\prime ,
\label{eq:tprimeapprox}
\end{equation}
when $T^\prime$ is small, which is the case in our experiments.

Previous imaging experiments by Anderson \textit{et al}\cite{Anderson2011} studied the transmission of light in a thin sample of DO11-doped PMMA as a function of time while the material is recovering. Assuming that the recovery process is due to diffusion, these images can be used to determine the fractional change in the concentration as a function of position and time. To do so, the Beer-Lambert law is inverted to give
\begin{equation}
\frac{C}{C_0} = 1-\frac{1}{\mathrm{abs}_0}\ln\left(\frac{T\left(x,t\right)}{T_B}\right) ,
\label{eq:concentrationBL}
\end{equation}
where $C_0$ is the initial concentration and $\mathrm{abs}_0$ is the absorption coefficient at the peak wavelength of the light source used to construct the image.  For a $100\,\mu$m thick sample at an initial concentration of $9\,$g/L and constant absorption cross section, $\mathrm{abs}_0 = 2.56$. $T$ is the transmittance of the sample, and $T_B$ is the baseline transmittance that is measured prior to illumination with the pump light that is used to damage the material.

\subsection{Parameters for pure diffusion}
\label{sec:fullrec}

\begin{figure}
\includegraphics[scale=1]{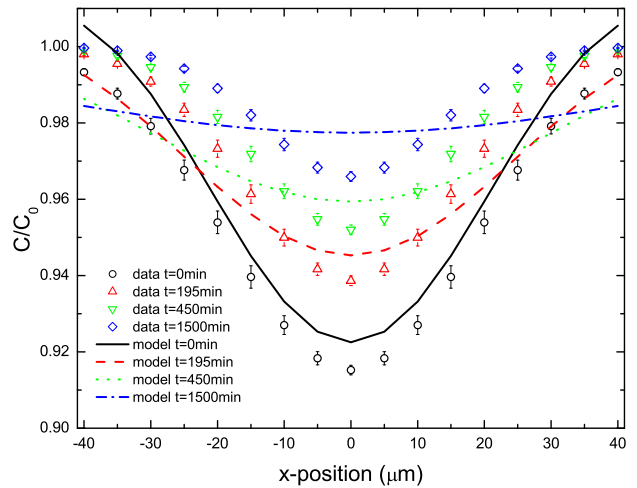}
\caption{The concentration fraction from the gaussian fits of experimental data(points) plotted along with the diffusion results(curves). We find that $\sigma = 24.4\,\mu$m$^{-1}$, $D = 0.767\,\mu$m$^2$min$^{-1}$, $MQ = 23.6\,\mu$m$^2$min$^{-1}$.}
\label{fig:purediff}
\end{figure}

The Gauss-Newton method was used to fit the output data generated by the Crank-Nicolson method to the concentration fraction determined from the experimental data. The elements of the Jacobian were estimated for the iterative fit of the partial differential equation.

Figure \ref{fig:purediff} shows a plot of the measured concentration fraction profile (symbols) and the concentration fraction calculated from Equation \ref{eq:diffTempxexp} (curves) as a function of time using as the initial condition the image of the damaged sample at the start of self healing. For pure diffusion, the concentration distribution broadens over time. However, the data clearly shows that only the amplitude changes, not the width.  Thus, we conclude that a pure diffusion mechanism cannot explain the data.

\subsection{Parameters for diffusion and photochemistry}
\label{sec:partrec}

We now consider the case where there are two processes: a permanent photochemical burn and dye diffusion. The permanent burn is identified by the residual change in transmittance at $1500\,$minutes. The image is found to be constant for up to $4500\,$minutes. Thus, we subtract this residual transmittance to isolate what we hypothesize is the diffusion component, which we analyze using Equation \ref{eq:concentrationBL}.

\begin{figure}
\includegraphics[scale=1]{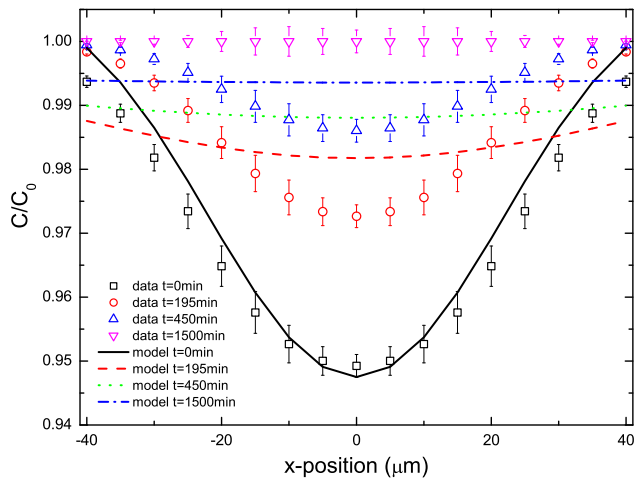}
\caption{The concentration fraction from the gaussian fits of experimental data plotted against the diffusion results. We find that $\sigma = 24.0\,\mu$m$^{-1}$, $D = 5.37\,\mu$m$^2$min$^{-1}$, $MQ = 18.5\,\mu$m$^2$min$^{-1}$.}
\label{fig:burndiff}
\end{figure}

The new base-lined data is shown in Figure \ref{fig:burndiff}, which is fit to the diffusion model. If the sample were to relax due to diffusion after the laser is turned off, the diffusion equation predicts that the width of the concentration profile increases. Again, we find that the data and theory disagree, so that diffusion does not appear to contribute.

\section{Conclusions}

We have considered two possible scenarios: pure diffusion and diffusion with permanent chemical changes due to photodegradation.  In both cases, the theory of diffusion and the experimental results do not agree. Thus, the contribution of diffusion to self healing of DO11 in PMMA polymer must be negligible.

The isobestic point and the evolution of the linear absorbance spectrum of DO11 in PMMA suggests that the damaged species is structurally distinct from the initial molecule.  While diffusion due to localized laser heating may be present, our measurements show that its effect is negligible. Ruling out dye diffusion will allow other mechanisms to be tested in the quest towards developing an understanding of the self-healing phenomena.

We would like to thank the Air Force (Grant No: FA9550-10-1-0286) for their generous support. We also thank Ben Anderson for supplying imaging data.

\bibliographystyle{osajnl}

\end{document}